# Information capacity and transmission are maximized in balanced cortical networks with neuronal avalanches

Woodrow L. Shew*[1], Hongdian Yang (杨黉典)*[1,2], Shan Yu (余山)[1], Rajarshi Roy[2], Dietmar Plenz[1]

[1] Section on Critical Brain Dynamics, Laboratory of Systems Neuroscience, National Institutes of Mental Health, Bethesda, MD, USA 20892;

[2] Institute for Physical Science and Technology, University of Maryland, College Park, MD, USA 20742

* These authors contributed equally to this work

**The repertoire of neural activity patterns that a cortical network can produce constrains the network's ability to transfer and process information. Here, we measured activity patterns obtained from multi-site local field potential (LFP) recordings in cortex cultures, urethane anesthetized rats, and awake macaque monkeys. First, we quantified the information capacity of the pattern repertoire of ongoing and stimulus-evoked activity using Shannon entropy. Next, we quantified the efficacy of information transmission between stimulus and response using mutual information. By systematically changing the ratio of excitation/inhibition (E/I) in vitro and in a network model, we discovered that both information capacity and information transmission are maximized at a particular intermediate E/I, at which ongoing activity emerges as neuronal avalanches. Next, we used our in vitro and model results to correctly predict in vivo information capacity and interactions between neuronal groups during ongoing activity. Close agreement between our experiments and model suggest that neuronal avalanches and peak information capacity arise due to criticality and are general properties of cortical networks with balanced E/I.**

**INTRODUCTION**
In the cortex, populations of neurons continuously receive input from upstream neurons, integrate it with their own ongoing activity, and generate output destined for downstream neurons. Such cortical information processing and transmission is limited by the repertoire of different activated configurations available to the population. The extent of this repertoire may be quantified by its entropy $H$; in the context of information theory, entropy characterizes the information capacity of the population (Shannon, 1948; Rieke et al., 1997; Dayan and Abbott, 2001). Information capacity is important because, as the name suggests, it defines upper limits on aspects of information processing. For example, consider the information transmitted from input to output by a population that only has two states in its repertoire ($H \leq 1$ bit). No matter how much information the input contains, the information content of its output cannot exceed 1 bit. A network with low entropy presents a bottleneck for information transmission in the cortex. Thus, it is important to understand the mechanisms that modulate the entropy of cortical networks.

Cortical activity depends on the ratio of fast excitatory (E) to inhibitory (I) synaptic inputs to neurons in the network. This E/I ratio remains fixed on average even during highly fluctuating activity levels (Shu et al., 2003; Okun and Lampl, 2008). However, it is currently not known whether a particular E/I ratio is advantageous for certain aspects of information processing. The existence of such an optimal ratio is suggested by two competing effects of E/I on entropy. First, a large E/I ratio, i.e. if excitation is insufficiently restrained by inhibition, can cause very high correlations between neurons (Dichter and Ayala, 1987). Since increased correlations decrease entropy (Rieke et al., 1997; Dayan and Abbot, 2001), we anticipate that a sufficiently large E/I ratio limits information transmission. This is consistent with findings that moderate levels of correlation can play an important role in population coding (Pola et al., 2003; Jacobs et al., 2009). At the other extreme, i.e. a small E/I ratio, weak excitatory drive is expected to reduce correlations as well as the overall level of neural activity. Although reduced correlations can lead to higher entropy, this increase may be counteracted by a concurrent drop in activity. Sufficiently suppressed activity reduces entropy (Rieke et al., 1997; Dayan and Abbot, 2001). Accordingly, we hypothesize that cortical entropy and information transmission are maximized for an intermediate E/I ratio.

Here we tested our hypothesis experimentally in cortex cultures, anesthetized rats, and awake monkeys and compared our results with predictions from a computational model. We discovered an optimal intermediate E/I ratio distinguished by 1) maximal entropy and 2) maximal information transmission between input and network output. This finding was based on analysis of both ongoing and stimulus-evoked population activity. Moreover, at this optimal E/I ratio, ongoing activity emerges in the form of neuronal avalanches (Beggs and Plenz, 2003) and interactions within the network are moderate. Agreement with our model suggests that by maintaining this particular E/I ratio the cortex operates near criticality and optimizes information processing.

**MATERIAL AND METHODS**

*In vitro MEA recordings and pharmacology.* Cortex-VTA (ventral tegmental area) organotypic co-cultures were grown on planar integrated multi-electrode arrays (MEA; for details see Shew et al., 2009). In brief, the MEA contained 60 recording electrodes (8x8 grid with no corner electrodes, 30 μm diameter, 200 μm inter-electrode spacing). LFP signals were sampled at 4 kHz and low-pass filtered (50 Hz cut off). Experiments consisted of a 1 hour recording of ongoing activity, followed by $0.5 - 1$ hr of stimulus-evoked activity. Next, the antagonists were bath-applied and recordings of ongoing activity followed by stimulus-evoked activity were repeated. We applied either 1) DNQX (0.5-1 μM) + AP5 (10-20 μM) or 2) Picrotoxin (PTX, 5 μM). The stimuli consisted of brief electric shocks delivered through a single electrode of the array located near layers 2/3 of the cortex culture. Each stimulus had a bipolar time course with amplitude –S for 50 μs in duration followed by an amplitude at +S/2 for 100 μs. We tested three similar sets of stimuli, each with 10 different stimulus levels, in μA (S1=10, 20, 30, 50, 60, 80, 100, 120, 150, 200; S2=6, 12, 24, 50, 65, 80, 100, 125, 150, 200; or S3=6, 12, 24, 50, 74, 100, 150, 200, 300, 400.) Different amplitudes were applied in pseudorandom order at 5 s intervals.

*MEA recordings in monkey.* All procedures were in accordance with NIH guidelines and were approved by the NIMH Animal Care and Use Committee. 96-channel MEA (10x10 grid with no corner electrodes, 400 μm separation, and 1.0 mm electrode length; BlackRock Microsystem) were chronically implanted in the right arm representation region of pre-motor cortex of two monkeys (Macaca mulatta, adults, one male, one female). Ongoing activity was recorded for 30 min. Monkeys were awake, but not engaged by any task or controlled sensory stimulation. Extracellular signals were sampled at 30 kHz and filtered off-line (1 to 100 Hz; phase neutral, 4[th] order Butterworth). For Figure 5B, a 4x4 subset of electrodes was analyzed, matching spatial dimensions of the 4x4 coarse resolution view of the *in vitro* data (see below). For Figure 5C, 4x4 patterns based on a coarse-binned set of 8x8 electrodes spanning larger spatial area were analyzed.

*MEA recordings in rats.* Urethane anesthetized rats aged $15 - 25$ days were studied. As described previously (Gireesh and Plenz, 2008), an MEA was inserted into superficial layers of barrel cortex (Michigan probe 8x4 electrodes, 200 μm interelectrode distance). Ongoing activity was recorded for $20 - 30$ min (4 kHz sampling rate, $1 - 3,000$ Hz bandwidth) and filtered between $1 - 100$ Hz (phase neutral, 4[th] order Butterworth) to obtain the LFP. In Figure 5B, the 8x4 MEA was coarse binned into a 4x4 pattern with 2 electrodes contributing to each bit. In Figure 5C, 4x4 patterns based on only 16 electrodes were also analyzed.

*Population event detection.* For each electrode, the standard deviation (SD) of the LFP was calculated. Automated (Matlab) population event detection entailed first identifying large ($< -4SD$ *in vitro*, -3SD *in vivo*) negative LFP fluctuations. Second, the time stamp $t_i$, amplitude $a_i$, and electrode $e_i$ of each negative peak (nLFP) occurring during these large fluctuations was extracted. Next, consecutive nLFPs were assigned to the same population event if they occurred sufficiently close ($t_{i+1} - t_i < \tau$) in time. The time threshold $\tau$ was determined based on the inter-nLFP-interval distribution, as described previously (Shew et al., 2009).

*Event size distribution, neuronal avalanches, and κ.* We defined the size *s* of a population event as the sum of all $a_i$ that occur during the event. During one recording we typically measured thousands of population events. We used a statistical measure called κ to quantify the character of each recording. The measure κ was developed to provide a network-level gauge of the E/I ratio, as demonstrated in a previous study (Shew et al., 2009). Similar to the Kolmogorov statistic, κ is computed based on the cumulative probability distribution of event sizes, *F(β)*, which describes the probability of observing an event with size less than *β*. More specifically, κ quantifies the similarity of the measured *F(β)* and a theoretical reference distribution $F^{NA}(β)$. The reference distribution was chosen based on empirical observations that cortical networks with unaltered E/I tend to generate a probability distribution of population event sizes, $P(s) \sim s^{-1.5}$, which defines neuronal avalanches (Beggs and Plenz, 2003; Shew et al., 2009). Importantly, population events are not considered to be neuronal avalanches unless they have this statistical property. Thus, $F^{NA}(β) \sim β^{-0.5}$ is the cumulative distribution corresponding to neuronal avalanches and κ assesses how close the observed distribution is to that of neuronal avalanches. By definition, κ is 1 plus the sum of 10 differences between *F(β)* and $F^{NA}(β)$.

$$\kappa = 1 + \frac{1}{10} \sum_{i=1}^{10} F^{NA}(\beta_i) - F(\beta_i) \, . \tag{1}$$

Therefore, *κ≈1* indicates a good match with the reference distribution and the existence of neuronal avalanches. Positive (negative) deviations away from *κ≈1* indicate a hyper- (hypo-) excitable network and the absence of neuronal avalanches. The 10 points at which the differences are computed are logarithmically spaced over the range of measured event sizes. It was previously shown that κ is not sensitive to changes in the number of differences (5 to 100 were tried; Shew et al., 2009).

*Binary patterns and entropy H.* Each population event is represented by an 8x8 binary pattern with one bit per recording electrode. A bit is set to 1 if the corresponding electrode is active during the event; otherwise it is set to 0. The entropy of this set of patterns is defined as

$$H = -\sum_{i=1}^{n} p_i \log_2 p_i \, , \tag{2}$$

where *n* is the number of unique binary patterns and $p_i$ is the probability that pattern *i* occurs. In Figure 2B, the black curve is based on the 8x8 binary patterns which represent the 60 electrodes of the MEA. We also studied entropy at different spatial resolutions by coarse-binning and for different spatial extents by using sub-regions of the MEA. The green curves in Figs. 2D, 3B, and 4A were obtained by reducing spatial resolution through coarse-binning of 8x8 patterns into square 4x4 patterns. Each bit in the 4x4 pattern was dependent on the state of 4 neighboring 2x2 electrode sets; if at least one electrode was active the bit was set to 1. Reduced spatial extent was tested with 16-bit patterns based on only 16 electrodes arranged in a 4x4 square (Fig. 2D). As pointed out in the results, these 4x4 patterns also reduce potential undersampling bias when compared to 8x8 patterns.

For stimulus-evoked activity, binary patterns were defined based on LFP activity measured during 500 ms following the stimulus. If the measured response at an electrode exceeded -8SD of the noise, then the corresponding bit was set to 1, otherwise it was set to 0. The stimulation electrode was always set to 0.

Note that the lack of corner electrodes on the MEA means that the corner bits of 8x8 patterns are always zero. This implies that the maximum entropy we could possibly record for 8x8 patterns is $2^{60}$ rather than $2^{64}$. For coarse-binned 4x4 patterns, the likelihood that the corner bits are active is slightly lower (about 25% lower). These effects are present for all E/I ratios examined. Therefore, they may affect the absolute values of entropy measurements, but they are not important for our conclusions, which are primarily based on changes in entropy. This is further confirmed by the robustness of our results to selecting 4x4 subregions from the center of the MEA for which corner electrodes are not missing (Fig. 2D).

The calculation of entropy entails estimating the occurrence probability for each pattern. Therefore, H generally will depend on number N of observed patterns unless N is so large that the probability of each pattern is well represented by the samples recorded. H will be underestimated for sufficiently small N but becomes independent of N for sufficiently large N. To estimate potential 'undersampling bias' we computed corrected values following the quadratic extrapolation method (e.g. Magri et al., 2009). First, we randomly selected a fraction $f$ of samples from the full set of $N$ patterns. We recomputed the entropy for fractions $f$=0.1 to 1 in steps of 0.1. We repeated this 10 times for each $f$. Next, we fit the average $H$ versus $f$ data with the following function:

$$H(f) = H_0 - \frac{a}{fN} - \frac{b}{(fN)^2} \tag{3}$$

The fit parameter $H_0$ is the estimated corrected value reported in the results section.

*Mutual information MI.* From a set of $N$ binary patterns, we defined a participation vector $q_i$ (length $N$) for each recording site $i$. $q_i(j) = 1$ or $0$ indicated that site $i$ was active or inactive during event $j$. The interaction between site $i$ and site $j$ was quantified by the mutual information (Rieke et al., 1997; Dayan and Abbott, 2001) of $q_i$ and $q_j$ defined as

$$MI(q_i;q_j) = \sum_{a\in 0,1} \sum_{b\in 0,1} p(q_i=a,q_j=b) \log_2 \left( \frac{p(q_i=a,q_j=b)}{p(q_i=a)p(q_j=b)} \right), \tag{4}$$

where $p(x)$ is the probability of $x$, $p(x,y)$ is the joint probability $x$ and $y$. $MI$ quantifies (in bits) the information shared by the two sites and provides similar information as a cross correlation (Supplementary Fig. S1). The $MI$ values reported in Results were averages over all pairs of sites,

$$MI = \frac{2}{M(M-1)} \sum_{i=1}^{M-1} \sum_{j=i+1}^{M} MI(q_i;q_j). \tag{5}$$

For the *in vitro* experiments, we also used mutual information in a different way to quantify the efficacy of information transmission between stimulus and response. Here we computed $MI(S;R) = H(R) - H(R|S)$. $H(R)$ is the entropy of the full set of response patterns for all stimuli. $H(R|S)$ is the conditional entropy, i.e. the response entropy for single stimuli, averaged across the different stimuli (Rieke et al., 1997; Dayan and Abbott, 2001).

*Likelihood of participation L* Likelihood of participation $L_i$ for site $i$ was defined as the fraction of patterns in which the site participated:

$$L_i = \frac{1}{N} \sum_{j=1}^{N} q_i(j) \,. \tag{6}$$

The average likelihood of participation $L$ for all $M$ sites is discussed in the main text and defined as

$$L = \frac{1}{NM} \sum_{i=1}^{M} \sum_{j=1}^{N} q_i(j) \,. \tag{7}$$

*Data shuffling to destroy interactions.* For the purpose of understanding how the entropy changes due to interactions between sites, we created surrogate data sets by shuffling the events in which sites participated. The 1s and 0s were randomly reordered in each participation vector $q_i$ such that interactions between sites were destroyed, but $L_i$ and $N$ remained fixed.

*Model.* The model consisted of $M$=16 binary sites (1=active, 0=inactive). Each site was intended to model the activity of a large group of neurons like the nLFP recorded at an electrode in the experiments. The strength of interactions between site $i$ and site $j$ was modeled as an activation probability; $p_{ij}$ was the probability that site $i$ would be activated at time $t+1$ if site $j$ was active at time $t$. Therefore, if a set $J(t)$ of sites were active at time $t$ then site $i$ would be active with probability $p_{iJ(t)} = 1 - \prod_{j \in J(t)} (1 - p_{ij})$. Increasing (decreasing) the E/I ratio was modeled by increasing (decreasing) the average $p_{ij}$ in the range between *0.1/M* and *1.5/M* (all $p_{ij}$ were initially drawn from a uniform distribution on [0,1] and then all were scaled down by dividing by a constant). Population events were modeled by activating a single initial site (like an electrical shock applied at a single electrode or a spontaneous activation in the experiments) and recording the resulting activations that propagated through the network. These dynamics were defined by

$$s_i(t+1) = \Theta\big[p_{iJ(t)} - \zeta_i(t)\big] = \Theta\Big[1 - \prod_{j \in J(t)} (1 - p_{ij}) - \zeta_i(t)\Big], \tag{8}$$

where $s_i(t)$ was the state of site $i$ at time $t$, $\Theta[x]$=0 (1) for $x<0$ (>0), and $\zeta_i(t)$ was a random number drawn from a uniform distribution on [0,1] at each update of each site. The states of all sites were updated simultaneously at each time step. Each population event in the model was represented with a 16 bit binary pattern (1= the site was active at least once during the response to the stimulus, otherwise 0). By simulating 1000 population events (always initiated at the same site), we generated a set of patterns for which the entropy was computed. From the event size distribution of network events we computed $\kappa$ (see Shew et al., 2009). The event size was defined as the sum of all activations from all sites during the population event. The range of average $p_{ij}$ values studied with the model resulted in a range of $\kappa$=0.6 to 1.6.

*Statistical Analysis.* For determining the statistical significance of differences in entropy for different drug conditions and differences in $\kappa$ for different drug conditions, we first used a one-way ANOVA to establish that at least one drug category was different from at least one other. Next we performed a post-hoc test of significant pairwise differences between the drug categories using a t-test with the Bonferroni correction for multiple comparisons. The same procedure was used to assess significance of differences in $H$ and $MI$ for different categories of $\kappa$.

## RESULTS

In all of our experiments, multi-electrode array (MEA) recordings of the local field potential (LFP, Fig. 1A) were used to obtain patterns of cortical population activity. We defined a recording site as 'active' if it presented a large, negative deflection in the LFP (Fig. 1A, green). We have previously demonstrated that such negative LFP deflections correlate with increased firing rates of the local neuronal population for each of the experimental preparations studied here: superficial layers of organotypic cultures (Shew et al., 2009), urethane anesthetized rat (Gireesh and Plenz, 2008) and awake monkeys (Petermann et al., 2009). We define a 'population event' as a set of electrodes which were active together within a short time (Methods). In our analysis, each population event was represented by a binary spatial pattern with one bit per recording site and 1 or 0 indicating an active or inactive site respectively (Fig. 1B; top). For each one hour recording *in vitro* (n=47) or 30 min recording *in vivo* (n=4 monkey, n=6 rat), we typically observed $10^3$ to $10^4$ population events.

First, we systematically explored a range of E/I conditions in cortex slice cultures. A reduced E/I ratio was obtained by bath-application of antagonists of AMPA- and NMDA-glutamate receptor mediated synaptic transmission (DNQX, 0.5-1 μM; AP5, 10-20 μM). This resulted in population events that were typically small in spatial extent (Figs. 1B and 1C, left). Conversely, an increased E/I ratio was obtained with an antagonist of fast GABA$_A$-receptor mediated synaptic inhibition (Picrotoxin; PTX, 5 μM), which led to stereotyped, spatially extended population events (Figs. 1B and 1C, right). In contrast, unperturbed E/I (Figs. 1B and 1C, middle) typically yielded a diverse pattern repertoire. The raster plots in Figure 1B (bottom) display examples of 100 consecutive population events recorded under the three different E/I conditions. Figure 1C displays example probability distributions of population event sizes for the three E/I conditions. We performed 11 recordings with reduced AP5/DNQX, 26 with no drug, and 9 with PTX. For each recording, we measured both ongoing activity and stimulus-evoked activity. For all recordings, we assessed the information capacity by computing the Shannon entropy of the full set of recorded binary patterns (Shannon, 1948; Rieke et al., 1997; Dayan and Abbott, 2001; see also Methods).

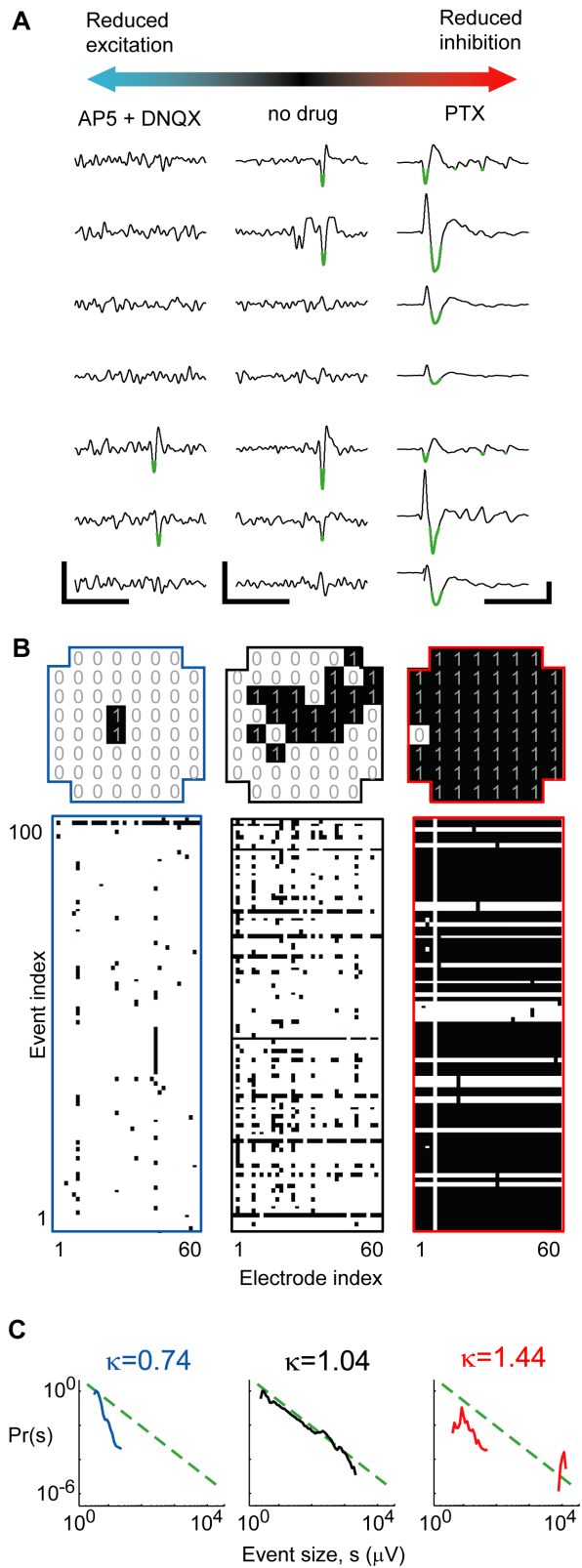

**Figure 1: Measuring the neural activation pattern repertoire for a range of E/I conditions. A** Example LFP recordings under conditions of reduced E (left), unperturbed E/I (middle), and suppressed reduced I (right). Scale bars: 250 ms x 10 μV (left, middle) and 250 ms x 100 μV (right). Population events were defined based on large negative deflections (<-4SD, green). **B** (top) Single examples of population events represented as binary patterns: 1=active site, 0=inactive. (bottom) Rasters including 100 consecutive population events represented as binary patterns; each row represents one event, each column represents one recording site. Left: reduced E. Middle: unperturbed. Right: reduced I. **C** Shape of event size distributions reveal changes in E/I, which are quantified with $\kappa$ (see methods; broken line: power law with exponent of -1.5).

**Peak information capacity of ongoing activity for intermediate E/I and neuronal avalanches.**

Our first finding was that the entropy $H$ for ongoing activity peaks at an intermediate E/I ratio. This was demonstrated with two different approaches. First, we compared entropy to the three pharmacological categories: AP5/DNQX, no drug, and PTX. We found that under the unperturbed E/I condition the average $H$ was significantly higher than either the reduced E/I condition of the AP5/DNQX or the increased E/I condition of PTX (Fig. 2A, ANOVA, $p$<0.05). Second, we compared entropy to a previously developed statistical measure called $\kappa$, which characterizes E/I based on population dynamics of the network (Shew et al., 2009; Methods). An advantage over the three pharmacology categories is that $\kappa$ is a graded measure, thus providing a continuous function of entropy $H$ versus E/I. A detailed definition of $\kappa$ is given in Methods. Briefly, $\kappa$ quantifies the shape of the population event size distribution, which is sensitive to changes in E/I (Fig. 1C): $\kappa < 1$ indicated reduced E/I and $\kappa > 1$ indicated increased E/I (Fig. 2B). Indeed, $\kappa$ was significantly different for the two pharmacological manipulations compared to the no-drug condition (Fig. 2B; p <0.05). When we plotted entropy versus $\kappa$ (Fig. 2C), we discovered a peaked function with maximum entropy occurring for $\kappa \approx 1$. This confirms our finding of peak entropy for the no drug condition (Fig. 2A) and provides a more refined view of the data; the peak occurred at $\kappa^* = 1.16 \pm 0.12$ (mean±SD, uncertainty determined by rebinning the experimental data, see Supplementary Fig. S2). The statistical significance of the peak in $H$ was confirmed by comparing $H$ for the ten experiments with $\kappa$ closest to 1 with the ten experiments with smallest $\kappa$ and ten with largest $\kappa$ ($p < 0.05$).

In addition to providing a graded measure of E/I, $\kappa$ assesses the statistical character of ongoing cortical population dynamics. Specifically, $\kappa \approx 1$ is the signature of neuronal avalanches (Shew et al., 2009), a type of population dynamics defined by a power-law population event size distribution with a power-law exponent near -1.5 (Beggs and Plenz, 2003; Stewart and Plenz, 2006; Gireesh and Plenz, 2008; Petermann et al., 2009; Shew et al., 2009). The computation of $\kappa$ entails first computing the difference between a measured event size distribution and a theoretical reference distribution defined as a power-law with exponent -1.5 (Fig. 1C, green dashed). Next, this difference is added to 1 (for historical reasons, Shew et al., 2009), resulting in $\kappa = 1$ for an exact match with a -1.5 power law, i.e. neuronal avalanches. In this context, our findings indicate that entropy is maximized under conditions which result in neuronal avalanches.

Next we tested the robustness of the peak in $H$ with respect to changes in spatial and temporal scales of recordings. First, as shown in Figure 2D (green), we found that the peak in $H$ remained close to $\kappa = 1$ ($\kappa^* = 1.01 \pm 0.02$), even when the original 8x8 patterns were coarse-grained to obtain 4x4 patterns at half the spatial resolution (see Methods). Second, the peak was also maintained when the spatial extent of the recorded area was reduced by 75% (4x4 electrodes near center of the MEA; Fig. 2D). Finally, we confirmed that the peak persisted for a restricted recording duration of 12 minutes rather than one hour (Fig. 2D, purple). The robustness of our finding to shorter recording durations is important since estimations of entropy depend on the number of samples recorded (see below).

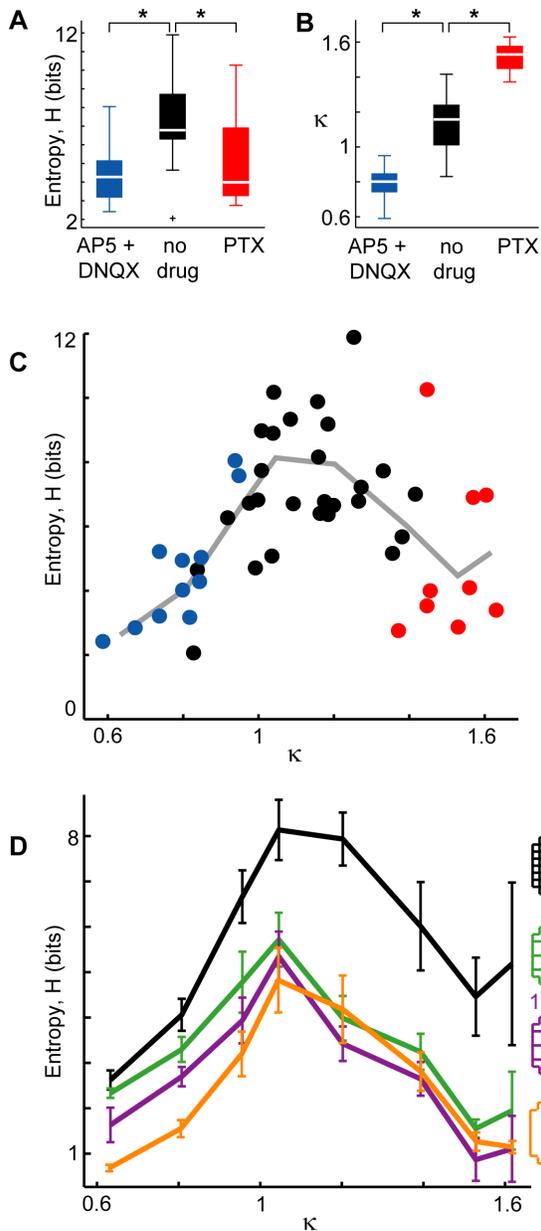

**Figure 2: Ongoing activity - peak information capacity at intermediate E/I ratio specified by $\kappa \approx 1$.** **A** Information capacity (entropy *H*) of the pattern repertoire is maximized for when no drugs perturb the E/I ratio. Significant differences marked with * (*p*<0.05). Box plot lines indicate lower quartile, median, upper quartile; whiskers indicate range of data, excluding outliers (+, >1.5 times the interquartile range). **B** The statistic $\kappa$, provides a graded measure of E/I condition based on network dynamics (methods). **C** Entropy *H* peaks near $\kappa \approx 1$. Each point represents one recording of ongoing activity (n=47, 8x8 MEA, 1 hr, color indicates drug condition; red=PTX, blue=AP5/DNQX, black=no drug). Line is the binned average of points. **D** The peak in entropy *H* is robust to changes in spatial resolution (green, 4x4 coarse-binned, 1 hr), spatial extent (orange, 4x4 subregion, 1 hr) and duration (purple, 4x4 coarse-binned, 12 min) of recording. (black, same data as in *A*). Error bars indicate mean±s.e.m.

**Peak information transmission between stimulus and response for intermediate E/I and neuronal avalanches.**

We now present measurements of stimulus-evoked activation patterns. A priori, one can expect a different distribution of stimulus-evoked patterns compared to ongoing activity and thus different entropy. Indeed studies suggest that ongoing activity is more diverse than typical stimulus-evoked activity (Welicky, 2004; Luczak et al., 2009; Churchland, 2010). However, if the entropy of evoked patterns changes with E/I in the same way that we found for ongoing activity, then evoked entropy may also peak near $\kappa$=1. This possibility is in line with significant evidence that ongoing activity in the cortex is intimately related to stimulus-evoked activity (Kenet et al., 2003; Ji and Wilson, 2007;

Han et al., 2008; Luczak et al., 2009). For instance, stimulus-evoked activity patterns recur during ongoing activity, both at the population level (Kenet et al., 2003; Han et al., 2008) and the level of spike sequences (Ji and Wilson, 2007) Therefore, our next aim was to test whether our finding of peak entropy near $\kappa$=1 also holds for stimulus-evoked activity.

Stimuli consisted of 10 different amplitude single bipolar shocks each applied 40 times in randomized order though a single electrode of the MEA within cortical layers II/III (Methods). A binary pattern was constructed to represent each response during the $20 - 500$ ms after the stimulus. The evoked entropy $H$ was calculated for the set of 400 stimulus-evoked activation patterns for each E/I. As found for ongoing activity, the evoked entropy was highest near $\kappa \approx 1$ for both fine and coarse spatial resolution (Fig. 3A; black - 8x8, green - coarse-grained 4x4, $p<0.05$).

In the introduction, we gave a simple example in which information transmission from input to output was limited due to low entropy. With our measurements of network responses (i.e. output) to stimuli (i.e. input), we can directly test whether efficacy of information transmission is optimized when entropy is maximized. This idea is concisely summarized in the following equation: $MI(S;R) = H(R) - H(R|S)$. Here, $MI(S;R)$ is the mutual information of stimulus and response which quantifies the information transmission (Rieke et al., 1997; Dayan and Abbott, 2001). $H(R)$ is the entropy of the full set of response patterns for all stimuli, while $H(R|S)$ is the conditional entropy, i.e. the average entropy per stimulus (Rieke et al., 1997; Dayan and Abbott, 2001). As shown above, $H(R)$ is maximized near $\kappa \approx 1$. Since, H(R|S) is always positive, $MI$(S;R) is bounded by H(R), and thus potentially also peaks near $\kappa = 1$. Indeed, we measured MI(S;R) under different E/I conditions and found that stimulus-response mutual information was maximized near $\kappa \approx 1$ (Fig. 3B; black - 8x8, green - coarse-grained 4x4, $p<0.05$).

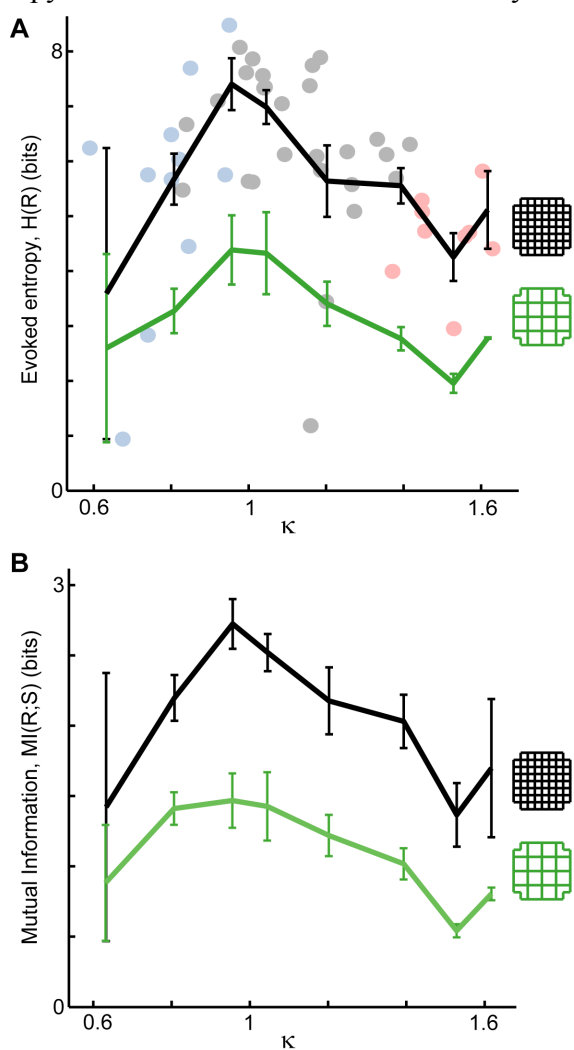

**Figure 3: Stimulus-evoked activity - peak information transmission at intermediate E/I ratio specified by $\kappa\approx1$. A** Single shock stimuli with 10 different amplitudes (10-200 μA) were applied 40 times each using a single electrode. The pattern repertoire of stimulus-evoked activity has maximum

entropy near $\kappa \approx 1$.  This holds for 8x8 response patterns (black line) as well as coarse resolution 4x4 patterns (green line).  Points correspond to 8x8 patterns: light blue – AP5/DNQX, gray – no drug, pink – PTX.  **B**  The efficacy of information transfer, i.e. mutual information of stimulus and response, also peaks near $\kappa \approx 1$.  The dashed line indicates the highest possible mutual information given 10 stimulus levels.  (black - 8x8; green – 4x4).  Error bars indicate s.e.m.

**Competition between activity rates and interactions explains peak in entropy.**

To identify and quantify the mechanisms leading to the peak in entropy near $\kappa=1$, we analyzed in more detail the coarse-grained 4x4 patterns measured during ongoing activity (Fig. 2D, green).  A priori, the total number of unique patterns that are possible is $2^{16}$, implying a maximum $H \leq log_2(2^{16})=16$ bits.  This maximum would be reached if all $2^{16}$ patterns occurred with equal probability.  However, during a 1 hr recording, the network did not generate all possible patterns, nor were different patterns equally likely, resulting in $H$ that was always below 16 bits.  The peak in $H$ was explained by three main factors that changed with the E/I ratio:  *i)* the number $N$ of patterns observed during the recording, *ii)* the likelihood $L$ that sites participate in patterns, and *iii)* the strength of interactions between sites.  The first two effects are related to the rates of observed activity and impose upper bounds on $H$:  effect *i* requires $H \leq log_2(N)$ (dash-dot line in Figs. 4A) and effect *ii* limits $H$ in a way that depends on $L$ (dashed line Figs. 4A).  Specifically, the highest possible entropy for a given $L$ can be computed by assuming that sites are independent,

$$H < -\sum_{i=1}^{M} \left( L_i \log_2 L_i + (1-L_i) \log_2 (1-L_i) \right),$$  (9)

where $M$ is the number of recording sites and $L_i$ is the likelihood of participation for site $i$.  This formula is based on the fact that the entropy of two independent systems combined is the sum of their individual entropies.  Since a single site $i$ is either active (with probability $L_i$) or inactive (with probability $1-L_i$), its entropy is $-L_i \log_2 L_i - (1-L_i) \log_2 (1-L_i)$.  Thus, adding the entropy of all sites, we obtain the formula above.  When $L<1/2$, increasing $L$ increases the upper bound on $H$.  When $L>1/2$, increasing L decreases the upper bound on H.  We found that $L$ increased over the range of E/I conditions we studied (Figs. 4C), while the number of patterns $N$ did not show a systematic trend.

We turn now to effect *iii*.  Increased interactions between sites always reduce $H$ due to the increased redundancy of the information at different sites (Schneidman et al., 2003).  We found that site-to-site interactions during ongoing activity increased with E/I (Figs. 4E), and quantified this trend in two ways.  First we computed mutual information (*MI*) between the activity recorded from different pairs of sites (Figs. 4E; red).  Note that above we used mutual information in a different way, computed between stimulus and response MI(R;S) to assess information transmission.  Second, we estimated the effect of interactions by computing the drop in entropy resulting from shuffling the data.  The shuffling procedure destroyed interactions by randomizing the set of population events in which each site participated, while keeping *L* and *N* fixed (Methods). The entropy of the shuffled data for the corresponding original *κ* value is shown in figure 4A (black) and, as expected, nearly reached the bounds set by the combined effects *i* and *ii*. The difference in entropy Δ*H* between the measured and shuffled data is due to interactions (Figs. 4E, blue). Δ*H* has previously been used to quantify redundancy (Dayan and Abbott, 2001).

In summary, at low E/I, effects *ii* and *iii* compete and effect *ii* wins, i.e. activity rates drop sufficiently low to cause low entropy even though interactions are also low. At high E/I, effects *ii* and *iii* cooperate, i.e. both high activity rates and strong interactions cause low entropy. Entropy peaked at an intermediate E/I ratio at which interactions between sites were not too low or too high (specified by *MI*≈0.2) and activity was not too depressed (*L*≈0.25).

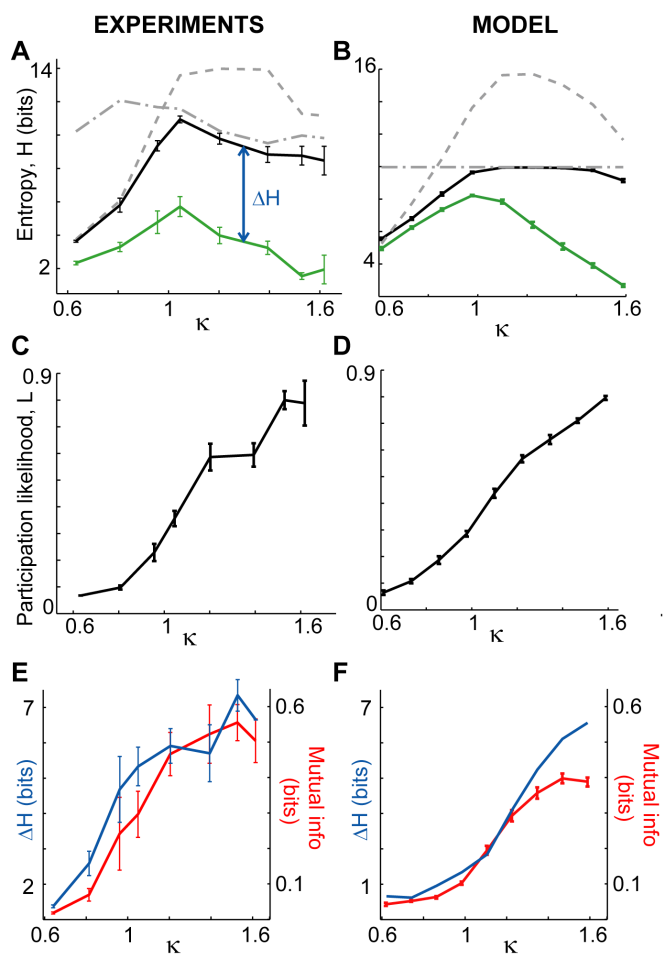

**Figure 4: Peak information capacity explained**. A detailed analysis of *in vitro* experimental results (left, Fig. 2b, green) and model results (right, Fig. 2c, blue) **A,B** Upper bounds on entropy are set by 1) the average likelihood that sites participate in patterns (dashed) and 2) the number of patterns observed (dash-dot). When the effects of interactions are removed by shuffling (methods), the entropy reaches these bounds (black), but the measured entropy (green) is always lower due to interactions. **C,D** Rise in participation likelihood L as E/I ratio is increased. This rise accounts for the bounds (dashed) shown in A,B. **E,F** Rise in interactions between sites (mutual information, red) is proportional to the loss in information capacity Δ*H* (blue). All error bars indicate s.e.m.

We remark that, if *N* were large enough (e.g. for longer recording duration), the upper bound due to effect *i* would become irrelevant, in which case, we still expect *H* to peak near *κ*=1 due to the combined effects of interactions (Δ*H*) and *L*. Nonetheless, the persistence of the peak in *H* for shorter duration recordings may be more relevant for cortex operations which occur on shorter time scales. We also tested the extent to which our measurements are impacted by sample size following the methods developed by Magri et al. (2009). The difference between our measured $H$ and 'corrected' $H$ was 0.06±0.06 (mean±SD) bits for 4x4 *in vitro* ongoing activity patterns and 0.22±0.18 bits for the 8x8 patterns. Thus, sample size effects are small compared to the variability from one experiment to another (see error bars in Fig. 2.) We also point out that $N$, $L$, and $MI$ are not the only factors that could potentially influence $H$. For example, not every site was equally likely to be active. Such spatial structure is expected to decrease entropy compared to a spatially homogeneous system with all other properties held fixed. This was not a major influence in our results.

**Experimental results confirmed in a computational network-level model.**

To gain further insight on our experiments, we compared our results to a network-level simulation, which has been used previously to model neuronal avalanches (Haldeman and Beggs, 2005; Kinouchi and Copelli, 2006; Shew et al., 2009). The model consisted of 16 binary sites. The state (1=active, 0=inactive) of each site was intended to represent a population of neurons in the vicinity of a recording electrode (Methods). The propagation of activity from one site to another was treated probabilistically; a connection matrix $p$ with entries $p_{ij}$ specified the probability that site $i$ would become activated due to site $j$ having been activated in the previous time step. Increases (decreases) in E/I were modeled by increasing (decreasing) the average $p_{ij}$ value through the range 0.006 to 0.1. For each 'E/I condition', 1000 population events were simulated, beginning with a single initially active site and the resulting patterns of activity were recorded. To facilitate comparison with our experimental results we also parameterized each E/I condition of the model using $\kappa$, based on population event size distributions.

In good agreement with our experiments, we found that entropy reached a peak for $\kappa \approx 1$ (Fig. 4B; green). Moreover, the explanation of peak entropy in terms of the competition between activity rates and site-to-site interactions also held for the model. Just as in the experiments, when the model data was shuffled to remove effects due to interactions, $H$ (Fig. 4B; black) approached the upper bounds set by the number of events (Fig. 4B; dash-dot) and the likelihoods of participation (Fig. 4B; dashed). The model $H$ results matched the experimental values, because the underlying changes in $L$ versus $\kappa$ (Fig. 4D) and the changes in site-to-site $MI$ versus $\kappa$ (Fig. 4F, red) were very similar to those measured experimentally. This agreement is not trivial; the same values of entropy could in principle be reached with different combinations of the underlying $L$ and $MI$ versus $\kappa$. For example, a peak in $H$ could result if $L$ remained fixed at 0.5 and interactions were minimized at $\kappa$=1. Site-to-site mutual information in the model reached slightly lower levels for high $\kappa$ when compared to experiments(Fig. 4F, red), which could be due to the lack of significant structure in the model connectivity matrix $p$.

### *In vivo* entropy matches *in vitro* prediction.

Finally, we analyzed recordings of ongoing activity from superficial cortical layers in two awake monkeys (premotor cortex) not engaged in any particular task and in urethane-anesthetized rats (n=6, barrel cortex) with no whisker stimulation. In agreement with previous studies (Gireesh and Plenz, 2008; Petermann et al., 2009), we found that the ongoing activity was organized as neuronal avalanches (Fig. 5A). More precisely, we found that κ=1.02±0.02 for the monkeys and κ=1.08±0.02 for the rats. Based on our *in vitro* findings, these κ values suggest that the *in vivo* networks are operating under E/I conditions that maximize entropy and information transmission. Although we cannot fully test this idea without a full range of κ *in vivo*, we can test whether the *in vivo* values of *H*, *L*, and *MI* match with those predicted from the *in vitro* results. As shown in Figure 5B and summarized in Table 1, we found good agreement with these predictions. We found no statistically significant difference between the *in vivo* results and the prediction from *in vitro* experiments with the same range of κ (1.0<κ<1.1, *p*<0.05). Nonetheless, the fact that entropy values *in vivo* were slightly higher than the *in vitro* results, may be due to the corresponding slightly lower *MI* values.

**Figure 5: *In vivo* properties predicted from *in vitro* results. A** Population event size distributions from ongoing activity in two awake monkeys (blue) and an example rat (green) are near a power-law with exponent -1.5 (dashed line), i.e. they exhibit neuronal avalanches and κ≈1. **B** In line with *in vitro* and model predictions for κ≈1, *in vivo* entropy

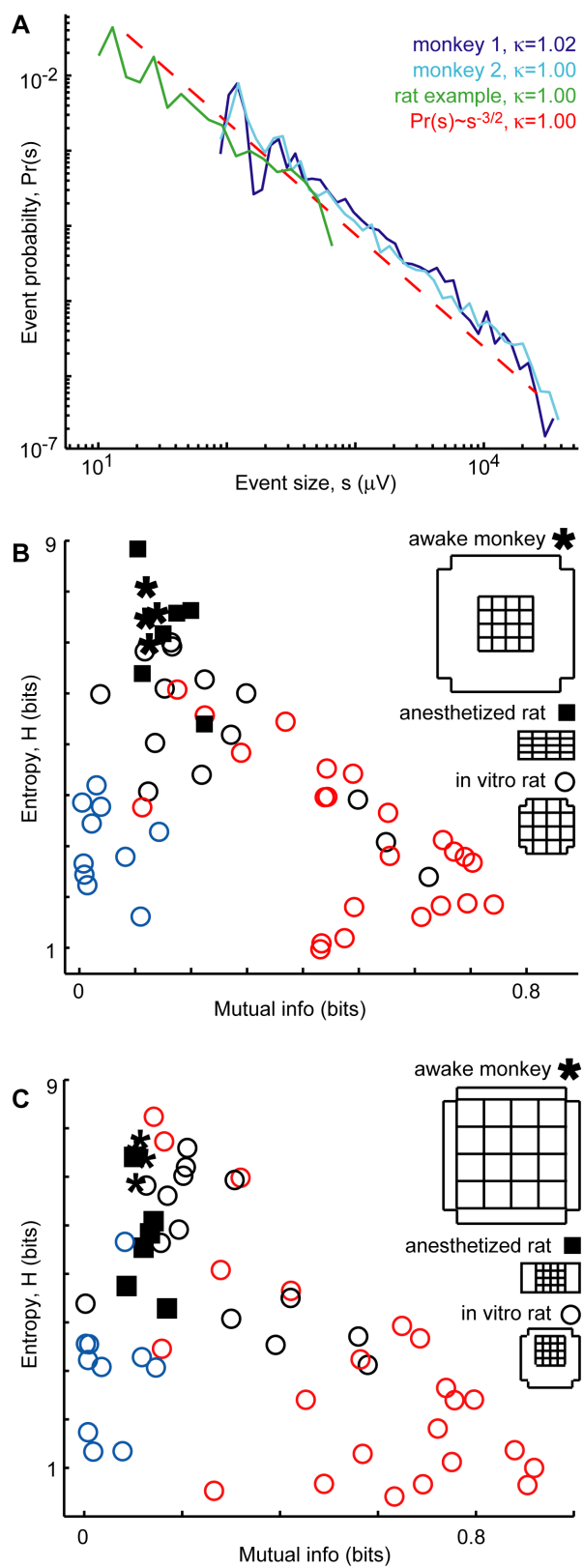

was high and mutual information between recording sites was moderate. (stars - two recordings on different days from each monkey; squares - anesthetized rats, n=6). The spatial extent of recorded area was approximately matched. **C** The result holds even when the spatial scales and resolution differ by factor of 4.

We note that the *in vivo* values of *MI*, which are based on LFP measurements, coexist with low values of pair-wise correlation *r* between spiking activity of units (mean±s.e.m. $r=0.03\pm0.01$, Supplementary Fig. S3), in line with recent reports for awake monkeys (Ecker et al., 2010) and anesthetized rats (Renart et al., 2010). The success of our prediction requires matching the number of recording sites (16 here), but is robust to large changes in spatial extent and resolution of recordings (Fig 5C). The prediction is also robust to changes in the threshold used for generating binary activity patterns from continuous LFP data (Supplementary Table S1).

| | Entropy, H (bits) | Participation likelihood, L | Site-to-site mutual information, MI (bits) |
|---|---|---|---|
| *In vitro* predictions for $1.0<\kappa<1.1$ | 5.7±1.6 | 0.3±0.1 | 0.2±0.2 |
| Awake monkeys $\kappa=1.02\pm0.02$ | 7.5±0.5 | 0.3±0.03 | 0.1±0.01 |
| Anesthetized rats $\kappa=1.08\pm0.02$ | 7.1±1.2 | 0.4±0.1 | 0.2±0.1 |

**Table 1:** In vivo results match in vitro predictions. Given the range of κ found in the in vivo recordings (1<κ<1.1), our in vitro results provide the predictions of H, L, and MI shown in the first row. The corresponding measurements from the awake monkeys (second row) and anesthetized rats (third row) match the in vitro predictions, i.e. they are not significantly different (p<0.05). Corresponding data are shown in Fig. 5B. All numbers are mean±SD.

## DISCUSSION

We employed in vitro and in vivo experiments as well as a computational model to study the effects of the E/I ratio on entropy and information transmission in cortical networks. We analyzed multisite measurements of LFP recorded during ongoing as well as stimulus-evoked activity. We found that entropy and information transmission are maximized for the particular E/I ratio specified by κ=1, which is the same E/I condition under which neuronal avalanches emerge.

We emphasize that the relative changes in *H* as we altered E/I are the meaningful results of our *in vitro* study; the absolute entropy values in bits depend upon arbitrary aspects of the analysis and measurements, e.g. the number of electrodes in the MEA. Thus, we are not suggesting that there is an absolute cap on the information that a cortical circuit can represent at ~10 bits and it is not appropriate to compare our $H$ values to those found in other studies of population entropy measures (e.g. Quian Quiroga and Panzeri, 2009). The important feature of our result is the *peak* in $H$ near $\kappa \approx 1$. We expect that any measure of population entropy would also peak for the same intermediate E/I, specified by $\kappa \approx 1$.

Previous studies have separately addressed the topics of entropy maximization (Laughlin, 1981; Dong and Atick, 1995; Dan et al., 1996; Li, 1996; Rieke et al., 1997; Dayan and Abbott, 2001; Garrigan et al., 2010), neuronal avalanches (Beggs and Plenz, 2003; Haldeman and Beggs, 2005; Stewart and Plenz, 2006; Ramo et al., 2007; Gireesh and Plenz, 2008; Tanaka et al., 2009; Petermann et al., 2009; Shew et al., 2009), and the balance of E/I (van Vreeswijk and Sompolinsky, 1996; Shadlen and Newsome, 1998; Shu et al., 2003; Okun and Lampl, 2008; Susillo and Abbott, 2009; Roudi and Latham, 2007), but our work is the first to show how these ideas converge in cortical dynamics.

Significant evidence suggests that maximization of entropy is an organizing principle of neural information processing systems. For example, single neurons in the blowfly visual system have been shown to exhibit spike trains with maximized entropy, considering the stimuli the fly encounters naturally (Laughlin, 1981). Applied at the level of neural populations, the principle of maximized entropy has provided successful predictions of receptive field properties in mammalian retina (Garrigan et al., 2010), lateral geniculate nucleus (Dong and Atick, 1995; Dan et al., 1996), and visual cortex (Li, 1996). Our work shows that the potential ability of a neural population in the cortex to achieve maximum entropy and maximum information transmission depends on the E/I ratio. Thus, if such properties are optimal for the organism, then the particular E/I ratio specified by $\kappa=1$ may best facilitate this goal.

We note that our investigation is not directly related to 'maximum entropy' models (e.g. Schneidman et al., 2006). In those studies, the aim was to use the maximum entropy principle (Jaynes, 1957) to find the simplest model to describe an experimental data set; entropy served as a modeling constraint. In contrast, here we compare the entropy across different experiments, searching for conditions which result in maximum entropy; entropy measurements are the results.

Several theory and modeling studies (including our own model) offer a deeper explanation of why $\kappa=1$ and neuronal avalanches occurs under E/I conditions which maximize entropy and information transmission (Beggs and Plenz, 2003; Haldeman and Beggs, 2005; Ramo et al., 2007; Tanaka et al., 2009). Recall that neuronal avalanches and $\kappa=1$, by definition, indicate a power-law event size distribution with exponent -3/2. This same property is found in many dynamical systems that operate near 'criticality'. Criticality refers to a particular mode of operation balanced at the boundary between order and disorder (e.g. Stanley, 1971; Jensen, 1998), akin to the balance of excitation and inhibition that we explore in our experiments. In our model, criticality occurs when the average $p_{ij}$ equals $1/M$ ($M$ is the number of sites). When $p_c > 1/M$, activity propagation is widespread and highly synchronous, like a seizure, while $p_c < 1/M$ results in weakly interacting, mostly independent neurons (Beggs and Plenz, 2003; Haldeman and Beggs, 2005; Kinouchi and Copelli, 2006). The balanced propagation that occurs at criticality might be attributed to interactions between excitatory and inhibitory neurons in the cortex. Using theory of Boolean networks, Ramo et al. (2007) showed theoretically that entropy of the event size distribution is maximized at criticality. Simulations of a model similar to our own showed that the number of activation patterns that repeat is maximized at criticality (Haldeman and Beggs, 2005). Tanaka et al. (2009) found that recurrent network models in which information transmission is optimized also exhibit neuronal avalanches and repeating activation patterns. Likewise, it has been shown that mutual information of input and output in feed-forward network models is maximized near criticality (Beggs and Plenz, 2003) and mutual information between neurons in the same network is maximized at criticality (Greenfield and Lecar, 2001). In line with these theory and model predictions, our results are the first experimental demonstration of peak entropy and information transmission in relation to criticality in the cortex.

Finally, a separate line of research has focused on the E/I ratio in cortical networks. Models emphasize the importance of balanced E/I for explaining the variability observed in spike trains (van Vreeswijk and Sompolinsky, 1996; Shadlen and Newsome, 1998), low correlations between spiking units (Renart et al., 2010), and generating diverse population activity patterns (Susillo and Abbott, 2009), which may play a role in memory (Roudi and Latham, 2007). Moreover, in vivo experiments have shown that synaptic input received by cortical neurons exhibits a fixed ratio of excitatory to inhibitory current amplitudes (Shu et al., 2003; Okun and Lampl, 2008). Since we measure $\kappa \approx 1$ in vivo, it follows that the 'balanced E/I' discussed in these previous studies may also correspond to the optimal E/I that we identify here.

In summary, our results suggest that by operating at the E/I ratio specified by $\kappa \approx 1$, the cortex maintains a moderate level of network-level activity and interactions which maximizes information capacity and transmission. This finding supports the hypotheses that balanced E/I and criticality optimize information processing in the cortex.

**Supplementary Material for 'Information capacity and transmission are maximized in balanced cortical networks with neuronal avalanches' by Shew et al.**

*Relationship between mutual information and correlation coefficient*
To quantify interactions between sites we used average pair-wise mutual information (MI). A more traditional approach is to use average pair-wise correlation coefficients (CC). Our reason for working with mutual information is two-fold. First, MI is less sensitive to noise when interactions are very weak. Second, MI arises from information theory and, thus, is a more natural fit with the study of entropy. Nonetheless, MI and CC are closely related. In Fig S1 we compare both quantities for the in vitro data presented in the main text. Also marked in Fig S1 are estimated theoretical bounds on the relationship between MI and CC. The lower bound is reached for L=0.5, while the upper bound corresponds to the extreme values of L=0 and L=1.

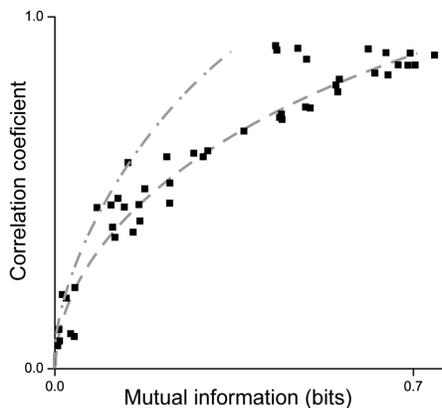

**Figure S1 Relationship between mutual information (MI) and correlation coefficient (CC).** The black points compare MI and CC for all in vitro experiments. The dashed line is an estimated lower bound, which is reached for L=0.5. The upper line is an estimated upper bound, which is reached for very low or very high L. The estimated bounds were obtained numerically for a single pair of binary vectors (10,000 events), each with the same L (0.1 to 0.9) and CC (0.1 to 0.9). These bounds are only approximate, because in the experiments, L is not the same from one electrode to another.

*Binning sensitivity for H vs. κ*

Here we show that the results shown in Fig 2C are robust to different choices of the bins used to produce the average line. The variability of the peaks of these curves was used to estimate the uncertainty in the conclusion that peak entropy occurs at κ≈1.

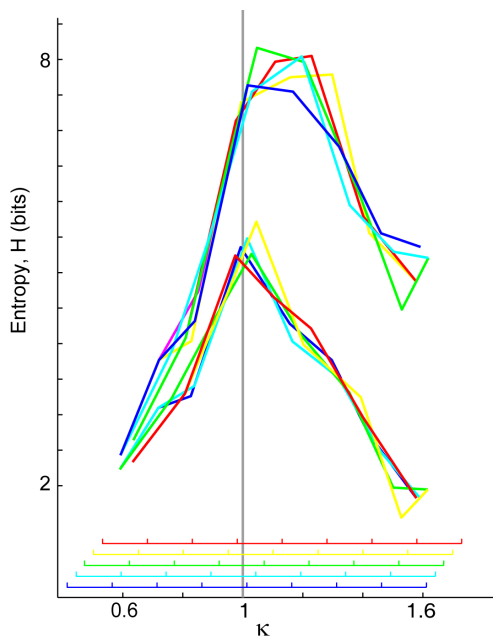

**Figure S2 Results robust to bin choices.** The results shown in Fig 2C (black, green) for H vs. κ for 8x8 patterns (top) and coarse-binned 4x4 patterns (bottom) were recomputed with different averaging bins. The different bin partitions are shown below the curves with corresponding colors.

*Pair-wise spike train cross correlation in the monkey*
Here we report the spike count cross correlation values between unit activity recorded simultaneously with the monkey LFP recordings discussed in the main text. The average correlation between unit signals is significantly lower than that between the population signals provided by the LFP. As shown in Fig S2, the mean, s.e.m., and distributions of correlation coefficients were in good agreement with recent reports from awake monkeys (Ecker et al., 2010) and anesthetized rats (Renart et al., 2010).
Spike sorting was performed with Plexon offline spike sorter (V2.8.8). 66 and 40 well isolated units were found for monkey 1 and monkey 2 respectively. Three principal components (PCA), peak-trough amplitude, and nonlinear energy were used as the sorting features. We defined 'well isolated' as follows: in a 2-D projection of at least 2 of the sorting features the unit must have a mean which is strongly different from the mean of noise waveforms (p≤0.001, multivariate ANOVA). If more than one unit was recorded from the same electrode, the difference between means of each unit was also required to be significant at this strict level (p≤0.001, multivariate ANOVA).
To compute spike count cross correlations between each pair of units recorded during ongoing activity we followed established methods (Renart et al., 2010). First, to obtain spike count vectors, the spike time stamps of each unit were 1) binned with 1 ms temporal resolution, 2) convolved with a Gaussian window with 50 ms width. The cross correlation coefficient was computed between all pairs (2145 pairs for monkey 1, 780 pairs for monkey 2) of spike count vectors.

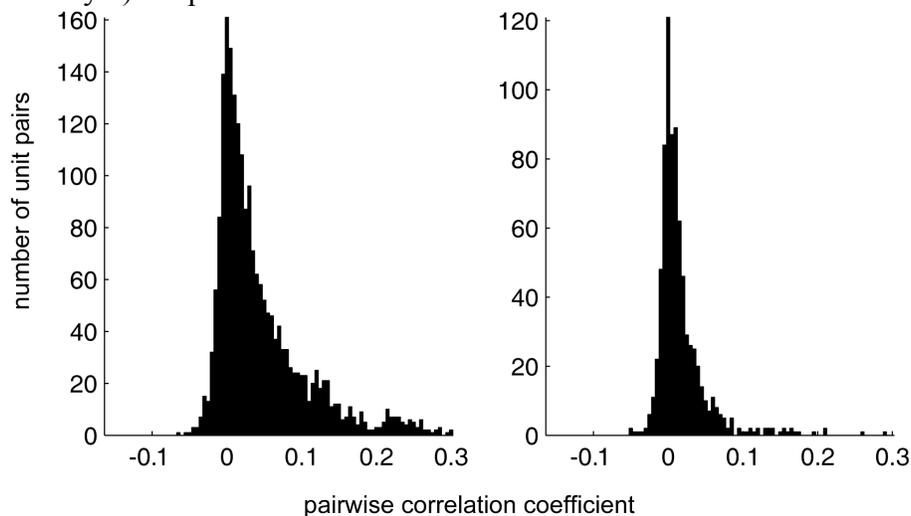

**Figure S3 Histograms of pairwise correlation coefficients of unit activity in the awake monkeys.** As reported previously (Ecker et al., 2010; Renart et al., 2010) the average across all pairs is near zero and positive. The histogram of all pairwise CC values for monkey 1 (left) and monkey 2 (right) are shown. The mean±sem CC values were 0.050±0.002 and 0.015±0.001 for monkey 1 and monkey 2 respectively.

*Robustness to event detection threshold*
In line with previous studies, our in vivo monkey results were robust to changes in the detection threshold for nLFPs. For thresholds -2.5, -3 and -3.5 SD we found no significant changes as shown in Table S1 below (mean±s.e.m.).

|  | κ | H | MI |
|---|---|---|---|
| *Monkey 1* | | | |
| *Day 1* | *1.02±0.02* | *7.96±0.35* | *0.11±0.01* |
| *Day 2* | *1.07±0.03* | *7.58±0.32* | *0.14±0.02* |
| *Monkey 2* | | | |
| *Day1* | *0.99±0.02* | *6.99±0.10* | *0.13±0.01* |
| *Day 2* | *1.00±0.02* | *7.51±0.08* | *0.11±0.02* |